\documentclass[twocolumn,showpacs,twoside,superscriptaddress]{revtex4}

\usepackage{amssymb, amsbsy, amsmath, latexsym, dsfont, array, layout, graphicx, mathrsfs}

\usepackage{color}

\newcommand{\one}{\mathds{1}}
\newcommand{\ket}[1]{\left|{#1}\right\rangle}
\newcommand{\bra}[1]{\left\langle{#1}\right|}

\begin{document}

\title{Quantum walk on a line for a trapped ion}
\author{Peng Xue}
\affiliation{Department of Physics, Southeast University, Nanjing 211189, P. R. China}
\affiliation{Institute for Quantum Information Science, University of Calgary, Alberta T2N 1N4, Canada}
\author{Barry C. Sanders}
\affiliation{Institute for Quantum Information Science, University of Calgary, Alberta T2N 1N4, Canada}
\author{Dietrich Leibfried}
\affiliation{National Institute of Standards and Technology, Boulder, CO 80305, USA}
\date{\today}

\begin{abstract}
We show that a multi-step quantum walk can be realized for a single trapped ion
with interpolation between quantum and random walk achieved by randomizing the
generalized Hadamard coin flip phase.
The signature of the quantum walk is manifested not only in the ion's position but also
its phonon number, which makes an ion trap implementation of the quantum walk feasible.
\end{abstract}


\maketitle
\emph{Introduction:--}
Quantum information processing promises revolutionary advances in
communication and computing with secure long-distance quantum key
distribution~\cite{GRTZ02} and quantum computing~\cite{Sho94} as two important long-term goals.
In the medium term, progress in quantum information implementations has been most
pronounced for quantum communication protocols~\cite{qcomm}, which consume entanglement to
enable quantum-enhanced communication.
An experimental quantum walk (QW)~\cite{Kem03,TM02,SBTK03} would be a major advance forward towards the ultimate goals of quantum information processing:
benchmarking coupling between qubit and `bus' mode,
test models of the environment and controlling decoherence,
and simulate exponentially-enhanced quantum algorithms over classical counterparts~\cite{CCD+03}.
Our aim is to realize a multi-step coined QW,
which would be implemented in an ion trap~\cite{exp}.

The random walk (RW) is ubiquitous in physics, chemistry, mathematics,
and computer science: this process underpins Brownian motion and
diffusion process, is used in satisfiability proofs (SAT), and is
intimately connected with the Wiener measure~\cite{CCD+03}. Quantization of
the RW~\cite{Kem03} has led to new quantum algorithms~\cite{AAKV01} and fascinating
physics such as decoherence-induced \emph{diffusion reduction}~\cite{SBTK03}. Our
goal is to see the QW realized in the laboratory. For this purpose, we
consider the simplest RW: a single walker whose two degrees of freedom
are position on regular one-dimensional lattice and a single two-sided
coin that generates random bits. Each coin flip generates a result
$0$ or~$1$, causing the walker to step left or right on the line,
respectively. In the quantum version of this walk on a line, each
position is a state in Hilbert space, and the coin is a qubit whose
flip is a unitary evolution: superpositions of position and
entanglement between the coin state and the walker's position are now
possible. In the QW, the walker commences at a point on the line and
alternates between unitary coin flipping and making left or right
steps that are entangled with the coin state.

This single-walker QW
has not yet been realized experimentally. Most proposals focus on
implementing the QW in phase space (walking around a circle in an
abstract position-momentum space)~\cite{SBTK03}, and even this easier case
(due to wandering in position space being strongly bounded) has not
been realized. Here we show that the QW on a line is indeed achievable
in an ion trap.

Travaglione and Milburn (TM) first proposed a QW implementation:
the walker's degree of freedom would be the position state~$|x\rangle$
of a single trapped ion~\cite{TM02}
and the coin state corresponds to the up state~$\left|\uparrow\right\rangle$ and down state~$\left|\downarrow\right\rangle$ of the ion's electronic degree of freedom.
Although a seminal proposal, unfortunately it is not viable, due mainly to four drawbacks:
(i)~impracticality of measuring~$x$;
(ii)~unavoidability of higher-order Lamb-Dicke (LD) contributions for large numbers of steps~\cite{CZ95};
(iii)~the need to control decoherence to enable interpolation between the RW and the QW
so that complementarity, hence quantumness, can be verified~\cite{KS05};
and most importantly (iv)~the impossibility of reading the coin state
(projection onto~$\left|\downarrow\right\rangle$ or~$\left|\uparrow\right\rangle$), because this procedure requires scattering many photons on the ion,
which inevitably destroys its motional state.
One suggestion for addressing the impracticality of~$x$-measurement by using
instead a quantum network and
multiple ions~\cite{FOBH05} is interesting but also impractical,
and suggestions of quantum walks on circles in phase space are also of value but
avoid entirely the core issue of realizing a QW on  a \emph{line}~\cite{TM02,SBTK03}.

TM's concept of an ion trap implementation of the QW is laudable,
but clearly major advances are required to bring their concept
of experimentally realizing a QW on a line to feasibility.
Here we overcome all four drawbacks:
(i)~we replace~$x$-measurement by measuring instead phonon number~$n$ via Rabi flopping and show that the phonon counts exhibit an unambiguous QW signature;
(ii)~we include higher-order LD contributions and show how they lead to breakdown of the QW on the line;
(iii)~we introduce a random-phase generalization to the coin flip and show how averaging enables interpolation between the QW and the RW~\cite{KS05}, and
(iv)~devise a method for inferring the phonon number distribution from electron shelving
by looking at the carrier transition instead of the first blue sideband~\cite{LBMW03}. In making the advance from concept to design of an ion-trap implementation for a QW,
we introduce some new and
valuable methods that have broad beneficial implications for ion-trap-based quantum
information processing in general.

\emph{Ideal QW on a line:--}
Before discussing the full, feasible implementation of the QW on a line in an ion trap,
here we study the ideal QW on a line and show that the phonon number distribution~$P_n$ carries the signature of
the QW as well as the inaccessible position distribution~$P(x)$.

Sufficient criteria for experimentally demonstrating the QW are:
a single walker whose position~$x$ is restricted to a one-dimensional lattice~$k\in\mathbb{Z}$
with step size~$\alpha$ between lattice points;
the position is incremented~$k\mapsto k+1$ if the coin state is~$\left|\uparrow\right\rangle$
and decremented~$k\mapsto k-1$ for~$\left|\downarrow\right\rangle$;
capability of unitarily flipping the coin such that $\left|\uparrow\right\rangle$,
$\left|\downarrow\right\rangle$ evolve to equal superpositions thereof;
quadratic enhancement of spreading for the QW compared to the RW; and
controllable decoherence to interpolate between the QW and the RW.

\begin{figure}[tbp]
   \includegraphics[width=8.5cm]{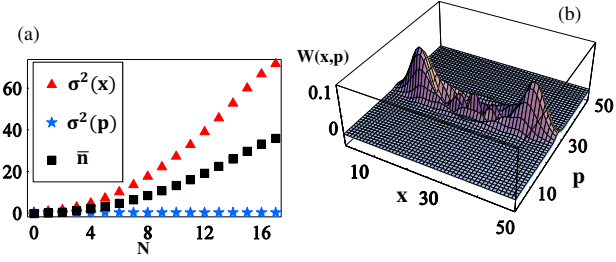}
   \caption{(a)~Mean phonon number and variances of position and momentum with~$\alpha=0.565$ up to~$N=17$ steps. (b)~Wigner function $W(x,p)$ for the walker state at~$N=10$.
   }
   \label{fig:idealwigner}
\end{figure}

Mathematically the unitary coin toss operator~$C$ is given by
$\sqrt{2}C(\phi)=\openone-\text{i}\hat{\sigma}_x\cos\phi+\text{i}\hat{\sigma}_y\sin\phi$
for $\openone$ the identity and
$\hat{\sigma}_x$ (flip), $\hat{\sigma}_y$ (phase-flip), and $\hat{\sigma}_z=\text{diag}(1,-1)$ (phase gate)
designating Pauli operators.
The phase~$\phi$ is arbitrary but must be constant for a unitarily evolving QW.
The walker's step to the left or right, entangled by the coin state, is enforced by the
unitary operator $T\equiv\exp\left(\text{i}\alpha\hat{p}\otimes\hat{\sigma}_z\right)$
for~$\hat{p}$ the position translation generator (i.e., a momentum operator)
and $\alpha$ the step size in a line.
Each step is effected by~$Q(\phi)\equiv T\left[\openone\otimes C(\phi)\right]$,
and the evolution to~$N$ steps is
$Q^N(\left\{\phi_\ell\right\})=\prod_{\ell=1}^N T\left[\openone\otimes C(\phi_\ell)\right]$.
For the ideal QW, $\phi_\ell$ is constant over all steps and typically assigned a value
of~$\pi/2$.
For~$\alpha_k\equiv(N-2k)\alpha$, an initial walker+coin state
$\left|\psi_0\right\rangle=\left|0\right\rangle(\left|\downarrow\right\rangle+\left|\uparrow\right\rangle)/\sqrt{2}$
evolves to
$\ket{\psi_N}
	=Q^N|\psi_0\rangle
	=\sum_{k=0}^{N-1}c^{(N)}_k\ket{\alpha_k,\downarrow }
		+\sum_{k=1}^N d^{(N)}_k\ket{\alpha_k, \uparrow}$, for
$\begin{pmatrix}c^{(N)}_k\\d^{(N)}_{k+1}\end{pmatrix}
		 =\frac{1}{\sqrt{2}}\begin{pmatrix}1&\text{i}\\\text{i}&1\end{pmatrix}\begin{pmatrix}c^{(N-1)}_k\\d^{(N-1)}_k\end{pmatrix}$.
The reduced walker state is $\rho_N\equiv\text{Tr}\left(\ket{\psi_N}\bra{\psi_N}\right)$,
which has position distribution $P(x)=\bra{x}\rho_N\ket{x}$
and phonon number distribution $P_n=\bra{n}\rho_N\ket{n}$.
Position variance~$\sigma^2(x)$, momentum variance $\sigma^2(p)$, and mean phonon number~$\bar{n}$ are shown in Fig.~\ref{fig:idealwigner}(a):
evidently $\sigma^2(x)\propto N^2$ and $\sigma^2(p)\sim$ constant up to $N=17$ steps;
for the RW, $\sigma^2(x)\propto N$, and this quadratic enhancement of position spreading is a signature
of the QW.

Fig.~\ref{fig:idealwigner}(a) also shows that the mean phonon number $\bar{n}\propto N^2$, which we identify as an alternative QW signature:
our choice of initial state yields
$	
	\bar{n}=\left[\sigma^2(x)+\sigma^2(p)\right]/2\propto N^2$,
whereas, for the RW,~$\bar{n}\propto N$.
Therefore, a quadratic enhancement in spreading of~$\bar{n}$ is just as good a signature of
the QW as is enhanced spreading of~$\sigma^2(x)$!
Fig.~\ref{fig:idealwigner}(b) reveals the invariance of the state under $x\mapsto -x$ or~$p\mapsto -p$ mappings
by depicting the Wigner function~\cite{W32} for the state after ten steps. Position~$x$ and momentum~$p$ distributions are marginal distributions of~$W(x,p)$

\emph{Controlling decoherence:--}
Although decoherence occurs naturally in the laboratory, for example due to magnetic field fluctuations,
which generate random $z$-rotations, this undesirable decoherence is largely eliminated by
spin echos (built into our technique).
To introduce controllable decoherence, we uniformly randomly choose each~$\phi_\ell\in(-\pi/q,\pi/q)$ with $q\geq 1$ a controllable parameter that
yields the RW for $q=1$ and the QW for $q\rightarrow\infty$.
The resultant walker state is
$\rho(\left\{\phi_\ell\right\})$.
Decoherence is achieved by running the experiment many times with different random
sequences~$(\left\{\phi_\ell\right\})$ each time, then averaging the density matrix at the $N^\text{th}$ step
of each run to obtain $\bar{\rho}_N$.
For~$\bar{\sigma}(x)$ the position-spread
and~$\bar{\bar{n}}$ the mean phonon number for~$\bar{\rho}_N$,
we conjecture the power-law rules
$\ln\bar{\sigma}(x)\propto\varsigma\ln N$ and $\ln \bar{\bar{n}}\propto\xi\ln N$,
with~$\left(\varsigma\approx 1/2,\xi\approx 1\right)$ for $q=1$ and
$\left(\varsigma\rightarrow 1,\xi\rightarrow 2\right)$ for $q\rightarrow\infty$.
Thus, $q$ controls decoherence and interpolates between the two extremes of QW and RW.

\emph{Ion trap implementation:--}
A single trapped ion (e.g., \ $^9\text{Be}^+$) is confined in a radio frequency (RF) ion trap.
Electronic (coin) and motional (walker) degrees of freedom
are coupled by `carrier' and `displacement' laser beams~\cite{LDM+03}.
The carrier-beam difference frequency is set to the frequency difference of the coin states $\{\ket{\downarrow},\ket{\uparrow}\}$.
The difference frequency~$\delta$ of the `displacement' Raman beams is close to the ion's motional mode frequency~$\omega_z$.

The initial state is prepared by laser-cooling the ion to the motional and electronic ground state,~$\left|0\right\rangle\left|\downarrow\right\rangle$,
then applying a~$\pi/2$ pulse, which creates an equal superposition of~$\left|\downarrow\right\rangle$ and~$\left|\uparrow\right\rangle$.
By applying displacement Raman beams, the interaction Hamiltonian is
$\hat{H}_\text{I}=\left[\text{e}^{-\text{i}(\delta t-\varphi)}D(\text{i}\eta\text{e}^{\text{i}\omega_z t})+\text{hc}\right]
		\Big(\Omega_\downarrow\left|\downarrow\right\rangle\bra{\downarrow}
		+\Omega_\uparrow\left|\uparrow\right\rangle\bra{\uparrow}\Big)$,
for~$D(\alpha)\equiv\exp(\alpha\hat{a}^\dagger-\alpha^*\hat{a})$ the unitary displacement operator and the carrier Rabi frequencies $\Omega_\uparrow=-\Omega_\downarrow/2$:
This interaction approximates the desired evolution.

We expand~$\hat{H}_\text{I}$ perturbatively in powers of the LD parameter~$\eta$;
this expansion is valid provided that~$\eta\left|\langle\hat{a}\rangle\right|$ and
$\left|\langle\hat{a}^2\rangle\right|-\left|\langle\hat{a}\rangle\right|^2
	=\sigma^2(x)+\sigma^2(p)+2\text{Cov}(x,p)$ are small, so~$\sigma^2(x)$, $\sigma^2(p)$ and~$\text{Cov}(x,p)$ are each small.
Thus,
\begin{align}
	\hat{H}_\text{I}
		\approx&\Big\{\text{e}^{-\text{i}(\delta t-\varphi)}
		\sum_{\ell=0}^3\frac{\left[\text{i}\eta\left(\hat{a}^\dagger \text{e}^{\text{i}\omega_z t}
			+\hat{a} \text{e}^{-\text{i}\omega_z t}\right)\right]^\ell}{\ell!}\nonumber\\
		 &+\text{hc}+O(\eta^4)\Big\}\left(\Omega_\downarrow\left|\downarrow\right\rangle\bra{\downarrow}
			+\Omega_\uparrow\left|\uparrow\right\rangle\bra{\uparrow}\right).
\end{align}
Evolution~$U=\exp \left[-\text{i}\int_0^t \hat{H}_\text{I}(t') \text{d}t'\right]$ over time~$t$ is approximately
\begin{align}
\label{eq:U}
	D(2\Omega_\uparrow\eta t) B(2\Omega_\uparrow\eta^3t)
	 U_\text{off}\left(2\Omega_\uparrow\right)\left|\downarrow\right\rangle\bra{\downarrow}\nonumber \\
	+ D(-\Omega_\uparrow\eta t) B(-\Omega_\uparrow\eta^3t)U_\text{off}(-\Omega_\uparrow)\left|\uparrow\right\rangle\bra{\uparrow}+O(\eta^4),
\end{align}
with each product of unitary operators in the sum comprising resonant unitary evolutions, whose exponents are linear in~$t$, and non-resonant evolution, with rapid terms such as~$\exp\left(\text{i}\omega_zt\right)$.

Evolution~(\ref{eq:U}) is dominated by resonant terms
of the Taylor expansion of~$\hat{H}_\text{I}$ such as
$B(\beta)=\text{e}^{\beta/6\left[(\hat{a}^\dagger)^2\hat{a}+\hat{a}^\dagger\hat{a}\hat{a}^\dagger
			 +\hat{a}(\hat{a}^\dagger)^2\right]-\text{hc}}=\text{e}^{\beta/2\hat{a}^\dagger(\hat{n}+1)-\text{hc}}$,
which arises from the $3^\text{rd}$-order resonant terms, and
\begin{align*}
	U_\text{off}\left(2\Omega_\uparrow\right)
		\approx D\left(-\text{i}\Omega_\uparrow\eta\frac{\text{e}^{2\text{i}\omega_zt}}{\omega_z}\right)
			S(2z) B\left(\text{i}\Omega_\uparrow\eta^3\frac{\text{e}^{2\text{i}\omega_zt}}{\omega_z}\right)\\
		\times \text{e}^{2\text{i}\Omega_\uparrow\left[-\eta^2\frac{\sin\omega_zt}{\omega_z}\hat{a}^\dagger\hat{a}+
		\left(\eta^3\frac{\text{e}^{4\text{i}\omega_zt}
	-2\text{e}^{2\text{i}\omega_zt}}{24\omega_z}\hat{a}^{\dagger 3}+\text{hc}\right)\right]}
\end{align*}
to $O(\eta^4)$.
Here $S(z)=\exp\left\{\frac{1}{2}z^*\hat{a}^2-\text{hc}\right\}$,
for $z=\Omega_\uparrow\eta^2\left(\frac{\text{e}^{\text{i}\omega_zt}}{\omega_z}+\frac{\text{e}^{3\text{i}\omega_zt}}{3\omega_z}\right)$,
is the squeezing operator, which arises from $2^\text{nd}$-order resonant terms
of the Taylor expansion for~$\hat{H}_\text{I}$.
Compared to resonant terms, the effect of~$U_\text{off}$ on~$\bar{n}$ and~$\sigma(x)$ is small.
We also neglect commutators arising from expanding
$\exp\left\{-\text{i}\int_0^t \hat{H}_\text{I}(t') \text{d}t'\right\}$ except for~$D$,~$B$ and~$S$, which are non-negligible.
We now clearly understand the small and large contributions to~$U$.

Without~$B$ and~$U_\text{off}$, Eq.~(\ref{eq:U}) is essentially a displacement of
$2\Omega_\uparrow\eta t$ or~$-\Omega_\uparrow\eta t$
if the coin is~$\downarrow$ or~$\uparrow$, respectively.
These asymmetric steps can be replaced by identical leftward and rightward steps
by alternating two $\pi$-pulses on the spins with two displacement steps $U$ and $U^\dagger$ to yield a desirable evolution from~$U$:
$U_\text{tot}=(\one\otimes X)U^\dagger(\one\otimes X) U (\one\otimes C)$, with
$X=\left|\downarrow\right\rangle\bra{\uparrow}+\left|\uparrow\right\rangle\bra{\downarrow}$.
Ignoring higher-order on-resonant term~$B$ and off-resonant term~$U_\text{off}$
yields $U_\text{tot}\approx D(3\Omega_\uparrow\eta t)\left|\downarrow\rangle\langle\downarrow\right|
+ D(-3\Omega_\uparrow\eta t)\left|\uparrow\rangle\langle\uparrow\right|$
to obtain a symmetric step size of~$\sim\pm3\Omega_\uparrow\eta t$.

\begin{figure}[tbp]
   \includegraphics[width=8.5cm]{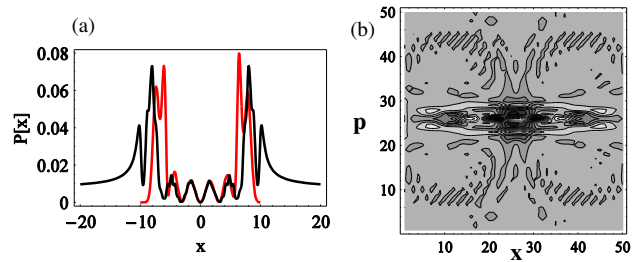}
   \caption{
   Quantum walker after $N=10$ steps.
   (a)~Position distribution for the ideal QW on a line (black) with~$\alpha=0.565 (\sim 3\Omega_\uparrow\eta t)$
  and for the ion trap implementation (red) with
  parameters~$(\delta,\omega_z,\Omega_\uparrow)/2\pi=(4,4,0.3)$MHz,~$\eta=0.1$ and
  pulse duration~$t=1\mu$s.
  (b)~Contour plot of Wigner function (min=$-0.095$, max=$0.052$)
  for ion-trap implementation, with each contour corresponding to
  a step of~$0.025$.
   }
   \label{fig:iontrapwigner}
\end{figure}

In Fig.~\ref{fig:iontrapwigner}(a) we see that the position peaks remain closer to the center compared to the ideal QW.
The change in displacement of peaks is small but nonnegligible as we can see from the following argument:
The evolution is dominated by displacements $D(3\Omega_\uparrow\eta t)\approx\one+3\Omega_\uparrow\eta t(\hat{a}^\dagger-\hat{a})$, with first-order term $\sim3\Omega_\uparrow\eta t\sqrt{\bar{n}}$
for $\left|\left\langle\hat{a}\right\rangle\right|=\sqrt{\bar{n}}=3N\Omega_\uparrow\eta t$,
whereas the $3^\text{rd}$-order~$B(\pm3\Omega_\uparrow\eta^3t)$ contribution
scales as~$\Omega_\uparrow t(\eta \sqrt{\bar{n}})^3/2$,
which is responsible for $\bar{n}$-dependent displacement.
Thus, the LD parameter needs to be kept small to ensure
the largest number of possible steps~$N_\text{max}$ with quadratic enhancement of spreading:
$\eta\ll \sqrt[4]{\frac{2}{3}}/\sqrt{N_\text{max}\Omega_\uparrow t}$.

Whereas higher-order resonant terms modify the QW as shown in Fig.~\ref{fig:iontrapwigner}(a),
non-resonant effects are much smaller than resonant contributions.
We test the contribution of each unitary operator in $U_\text{off}$ by forcing some unitary
operators to be identities; then we evaluate how much each of those operators affects the dynamics.
By this procedure we ascertain that non-resonant operator contributions $D\left(\mp\text{i}\Omega_\uparrow\eta\frac{\text{e}^{2\text{i}\omega_zt}}{\omega_z}\right)$ and $B\left(\pm\text{i}\Omega_\uparrow\eta^3\frac{\text{e}^{2\text{i}\omega_zt}}{\omega_z}\right)$ are primarily responsible for creating the pincer-like momentum sidebands
observed in the walker's Wigner function shown in Fig.~\ref{fig:iontrapwigner}(b).
These momentum sidebands are directly responsible for the small monotonic increase of momentum variance
observed in Fig.~\ref{fig:idealwigner}(a), whereas momentum variance is constant for the ideal QW on a line.

We establish numerically that squeezing~$S$ in the evolution
is responsible for slight asymmetry of the position distribution in Fig.~\ref{fig:iontrapwigner}(a).
This asymmetry arises because the position distribution peaks are squeezed conditioned on the coin state;
when probability amplitudes are coherently added,
the position distribution symmetry is slightly violated.
Off-resonant contributions also effect a small rotation of the Wigner function
in phase space due to a~$e^{\text{i}\theta\hat{n}}$ contribution to~$U_\text{tot}$ for
small~$\theta$.

\emph{Counting phonons:--}
Previously~\cite{LBMW03} the motional number distribution has been determined by driving the ion on the first blue sideband and Fourier transforming the atomic population
in the $\left|\downarrow\right\rangle$ as a function of drive duration
$P_{\downarrow}(t)=1/2 \Big[ 1+\sum_{n=0}^{\infty} P_n \cos(\Omega_{n,n+1} t)\Big]$,
for~$P_n$ the $n$-phonon probability
and $\Omega_{n,n+1}$ the $\ket{\downarrow, n}\leftrightarrow\ket{\uparrow, n+1}$ Rabi frequency.

Outside the LD regime,
$\Omega_{n,n+1}\propto \frac{\eta}{\sqrt{n+1}}L_n^1(\eta^2)$,
where the generalized Laguerre polynomial~$L_n^m(x)$ is nonmonotonic in~$n$
thereby leading to ambiguities in determining $P_n$ by this method.
This problem can be redressed for low~$\eta$
by Fourier transforming the carrier signal for the transition
$\ket{\downarrow, n}\leftrightarrow\ket{\uparrow, n}$) where
$\Omega_{n,n}\propto L_n^0(\eta^2)$~\cite{LBMW03}.
These frequencies $\Omega_{n,n}$ are monotonous and distinguishable for $n<60$,
which is a sufficient range to observe the hallmarks of QWs vs RWs.
For $\eta\leq 0.2$, the carrier alone is sufficient to find $P_n$ for $n<25$. However, once $P_n$ is known for $n<25$ this information can serve to lift the ambiguities on the blue sideband, which then in turn can be used to determine $P_n$ for $25\leq~n<60$.

\begin{figure}[tbp]
   \includegraphics[width=8.5cm]{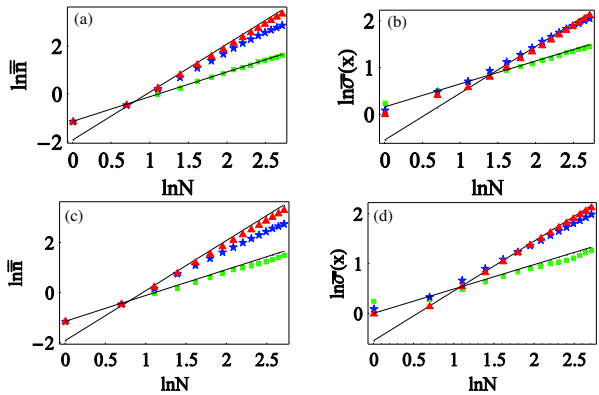}
   \caption{
        (color online)
        ln-ln plot of ideal walker's (a)~mean phonon number~$\bar{\bar{n}}$ and
        (b)~position spread~$\bar{\sigma}(x)$ and ion trap realization
of walker's
        (c)~mean phonon number and (d)~position spread
        as a function of step number~$N$
        for decoherence $q=1$ (green), $q=5$ (blue), and $q=20$ (red).
        Experimental parameters and $\alpha$ for the ion trap realization
(c, d) are as in Fig.~2.
        The slopes of the dotted lines, which interpolate from the ideal
        QW to RW (shown in each figure as solid lines with greatest and least
        slope, respectively), are
        (a)~$1.003$, $1.660$, and $1.938$,
        (b)~$0.510$, $0.860$, $0.990$,
        (c)~$1.061$, $1.530$, and $1.902$, and
        (d)~$0.496$, $0.824$, and $0.985$.}
   \label{fig:idealln}
\end{figure}

\emph{Simulations:--}
We simulate the walker+coin dynamics and calculate the mean phonon number~$\bar {\bar{n}}$
and position spread~$\bar{\sigma} (x)$ in two cases: the ideal walk on a line in Figs.~\ref{fig:idealln}(a, b)
and the walker as a trapped ion in Figs.~\ref{fig:idealln}(c, d).
Decoherence is controlled by~$q$: RW for $q\rightarrow 1$; QW for $q\rightarrow\infty$.
Step size $\alpha=0.565$ corresponds to $3\Omega_\uparrow\eta t$.

The ln-ln plots reveal the small- and large-$q$ power law relationship between
either $\bar{\bar n}$ or $\bar{\sigma}(x)$ and step number~$N$, as predicted;
furthermore the slopes approach~$1$ for the RW and~$2$ for the QW,
thus confirming that both $\bar{\bar{n}}$ and $\bar{\sigma}(x)$ suffice for observing
the RW-QW transition through controlling~$q$.
Slopes of $\bar {\bar{n}}$ and $\bar{\sigma}(x)$ for the ion trap case are slightly smaller than for the ideal QW, but this small degradation is fully explained,
namely nonlinear and nonresonant contributions to the evolution.
The RW-QW transition is excellent despite these pragmatic considerations.

\emph{Conclusions:--}
In summary we have developed a scheme for realizing the first single-walker QW in the laboratory,
with the ion's electronic degree of freedom serving as the two-state coin and the motion
as the walker's degree of freedom.
In contrast to current approaches to developing QW implementations,
which would realize QWs on circles in phase space~\cite{TM02,SBTK03},
our approach yields a RW-QW transition in position space.
In other words, the walker is truly spreading out over unbounded position space
rather than being folded back on itself.
Our approach is true to the spirit of RWs over unbounded domains but required
important innovations taking this idea well beyond TM's first concept for the QW in
an ion trap~\cite{TM02}.

Although the walk is over position, we show that the experimentally accessible phonon number
equally reveals the RW-QW transition.
We have shown that phonon number measurement is feasible
for dozens of phonons by driving the ions at the carrier frequency,
then Fourier transforming the ground state population to reveal the Rabi
frequencies~$\Omega_{n,n}$, hence the phonon number distribution.
This approach is similar to the approach of blue-sideband driving~\cite{LBMW03}
but is more effective in revealing $P_n$ over wide-ranging phonon number~$n$.
In addition, we introduce an experimentally controllable
phase randomization procedure that is parameterized by~$q$.
The RW-QW transition is a key part of any experiment
that plans to demonstrate QW behavior~\cite{KS05},
yet the ion trap dynamics are almost perfectly coherent.
Finally we have been quite careful in studying LD corrections, which are clearly nonnegligible.
In conclusion our theory establishes a pathway to realizing a many-step QW,
and our techniques for counting phonons
should be useful for general quantum information protocols.

\emph{Note:--}
Subsequent to submitting this manuscript, a trapped-ion three-step coined quantum walk,
which shows beautifully the difference between the QW and RW, has been reported~\cite{SMS+09}.
However, their walk is limited to three steps to avoid higher LD contributions, and
they measure motional wavepacket overlap rather than position.
Our scheme overcomes such limitations.

\emph{Acknowledgements:--}
This work was supported by Southeast University Startup fund, NSERC, MITACS, QuantumWorks, CIFAR, \emph{i}CORE, IARPA and the NIST quantum information program.

\end{document}